\documentclass[a4paper,12pt]{article}
\usepackage{mathptmx}
\usepackage{epsfig}
\usepackage{float}
\usepackage{amssymb}
\usepackage{latexsym}
\usepackage{amsfonts}
\newcommand{\beq}{\begin{equation}}
\newcommand{\eeq}{\end{equation}}
\newcommand{\beqa}{\begin{eqnarray}}
\newcommand{\eeqa}{\end{eqnarray}}

\def \h {\frac{1}{2}}

\def \r {\rho}
\def \t {\tau}
\def \D {D}

\def \R {\mathbb R}
\begin{document}\large
\title{\bf Application of \textit{afxy-code} for parameterization of ionization yield function Y in
the atmosphere for primary cosmic ray protons }

\author{L. Alexandrov and A. Mishev \\ \\
{\it Laboratory of Theoretical Physics, JINR,} \\
{\it 141980 Dubna, Russia}  \\
{\it and} \\
{\it Institute for Nuclear Research and Nuclear Energy-Bulgarian Academy of Sciences}\\
{\it Tsarigradsko Chaussee 72, 1784 Sofia, Bulgaria}
\\ \\
}

\maketitle
\begin{abstract}
In this work is obtained new approximation for the yield function Y
for cosmic ray induced ionization in the Earth atmosphere on the
basis of simulated data. The parameterization is obtained using
inverse nonlinear problem solution with afxy(analyze fx=y)-code.
Short description of the methods is given. The found approximation
is for primary proton nuclei, inclined up to 70 degrees zenith
angle. This permits to estimate the direct ionization by primary
cosmic rays explicitly. The parameterization is applicable to the
entire atmosphere, from ground level to upper atmosphere. Several
implications of the found parameterization are discussed.
\end{abstract}

 \normalsize

\label{cor}{\small Corresponding author:A.Mishev, INRNE-BAS, Tel:
(+359) 29746310; e-mail: mishev@inrne.bas.bg}

\section{Introduction}
Presently it is known that the Earth is hit by elementary particles
and atomic nuclei of very large energies and in wide energy range,
this is the cosmic ray radiation. The fluxes variate, from $10^4$
$m^{-2}$ $s^{-1}$ at energies $~10^{9}$ eV to $10^{-2}$ $km^{-2}$
$yr^{-1}$ at energies $~10^{20}$ eV. The cosmic ray intensity is
approximatively expressed with (1), where E is the total particle
energy per nucleon in GeV and $\alpha = -2.7$ is the spectral index.
The majority of these particles are protons. The primary cosmic rays
penetrate upper atmosphere, secondary cosmic rays are produced by
the interactions of primary rays in the atmosphere.

\begin{equation}
I_{n}(E)\varpropto1.8(\frac{E}{GeV})^{\alpha} nucleons.  cm^{-2}.
s^{-1}. sr^{-1}
\end{equation}

The abundances are approximately independent of energy, at least
over the dominant energy range of 10 MeV/nucleon through several
GeV/nucleon. By mass about 79 percent of nucleons in cosmic rays are
free protons, and about 80 percent of the remaining nucleons are
bound in helium.

There are three different types of cosmic rays: galactic cosmic
rays, solar cosmic rays and anomalous cosmic rays. Most galactic
cosmic rays are accelerated in the shock waves of supernova
remnants. Because of their deflection by magnetic fields, galactic
cosmic rays follow convoluted paths and arrive at the top of the
Earth's atmosphere in practice uniformly from all directions.
Galactic cosmic rays are the most typical cosmic rays, and their
flux in the solar system is modulated by the solar activity:
enhanced solar wind shields the system from these particles
\cite{For}.

We  know that the galactic cosmic rays create the ionization in the
stratosphere and troposphere and also in the independent ionosphere
layer at altitudes 50-80 km in the D region \cite{Vel74}. This
ionization is a result from the impact of the secondary cosmic ray
electromagnetic, muon and hadronic components on the planetary
atmosphere. The cosmic ray induced ionization is an important factor
of space influences on atmospheric properties.

In general the variations of cosmic ray induced ionization are
caused by solar activity variations, which modulate the cosmic ray
flux in interplanetary space, and changes of the geomagnetic field,
which affects the cosmic ray penetration in the atmosphere and their
access to Earth. The changing solar activity is responsible for the
variation of solar wind, respectively cosmic rays. The solar wind
reduces the flux of cosmic ray reaching the Earth, since a larger
amount of energy is lost as they propagate up the solar wind. Since
cosmic rays dominate the ionization, an increased solar activity
will translate into a reduced ionization.

 On the basis of satellite
data analysis was shown that cloud cover varies with the variable
cosmic ray flux reaching the Earth \cite{Sve97,Tin96}. Over the
relevant time scale, the largest variations arise from the 11-yr
solar cycle, and indeed, this cloud cover seemed to follow the cycle
and a half of cosmic ray flux modulation. Specifically was shown
that the correlation is primarily with low altitude cloud cover
\cite{Mar}.

 In this connection a detailed model of the cosmic ray
induced ionization will be a good basis for a quantitative study of
different mechanisms affecting Earth's atmosphere. To estimate the
cosmic ray induced ionization it is possible to use a model based on
an analytical approximation of the atmospheric cascade \cite{Kar} or
on a Monte Carlo simulation of the atmospheric cascade \cite{Uso04,
Des}. A key issue, which allows to estimate the cosmic ray induced
ionization for given location, altitude and spectrum of cosmic rays
is the use of ionization yield function Y (2) which is defined
according \cite{Uso06}

\begin{equation}
Y(x,E)=\triangle E(x,E)\frac{1}{\triangle x}\frac{1}{E_{ion}}\Omega
\end{equation}

where $\triangle E$ is the deposited energy in layer $\triangle x$
in the atmosphere and $\Omega$ is a geometry factor, integration
over the solid angle with zenith of 70 degrees. Afterwards the ion
pair production q by cosmic rays following steep spectrum is easy
calculated according the expression:

\begin{equation}
q(h,\lambda_{m})=\int D(E,\lambda_{m})Y(h,E)\rho(h)dE
\end{equation}

where $D(E,\lambda_{m})$ is the differential primary cosmic ray
spectrum at given geomagnetic latitude, Y is the yield function
according (2), $\rho(h)$ is the atmospheric density in
[$g.cm^{-3}$].

The ionization yield function Y depends only of assumed physical
models for cascade processes in the atmosphere, the atmospheric
model and the type of the primary particle. Therefore having a
convenient parameterization for ionization yield function Y and
parameterized differential energy spectrum of galactic cosmic rays
at the Earth's orbit it is possible to estimate cosmic ray induced
ionization in different locations and conditions. In this work we
use the previously obtained ionization yield function Y \cite{Mis07,
Vel07} for primary protons.

 The major contribution of cosmic ray
induced ionization in the atmosphere is due to particles having
energy till 1 TeV, taking into account the steep spectrum of primary
cosmic rays (1). This is the reason to deal in this paper with
primary protons with maximal energy of 1 TeV.

\section{How to investigate a given nonlinear system}
Many problems in physics, applied mathematics even and pure
mathematics lead to solution of nonlinear system of equation.

To analyze nonlinear equation, we write it in the form
\beq\label{vector} fx=\overline{y}, \quad f:\D_f \subset
\R^n\to\R^m,\; x\in\R^n,\;\overline{y}\in\R^m.\eeq

To analyze iteratively  nonlinear systems involves two related
classes of problems \cite{Ale07}- heuristic investigation of
nonlinear systems issuing from not-refined mathematical models and
automatic solution of streams of one-type nonlinear systems of
equations.

\subsection{Main iteration procedure in
\textit{afxy}-code\cite{Ale07}}

A powerfull tool for an heuristic analysis of systems of  nonlinear
equations (4) is \textit{afxy}-code (analyze fx=y).In the case
$n\leq m$ as well in non-trivial case $n>m$ we employ regularized
Gauss-Newton type iterator
  \beqa\label{GN}\overline{x}\in\D_f,\;
x^{0}\in \D_f, \; \epsilon^{0}>0,\quad ({f'}^T(x^k) f'(x^k)
+(\epsilon^{k}+\alpha^{k}) I)
(x^{k+1}-x^k)=\\
-{f'}^T(x^k)(fx-\overline{y})-\alpha^{k}\overline{x}),
\;k=0,1,\ldots\;. \nonumber \eeqa

The regularizators  $\varepsilon^{k}$ (of the process) and
$\alpha^{k}$ (of the problem) will be defined below.

In the right hand of (5) original problem (4) is transformed by
Gauss manner $Fx:=f^{'T}(x)(fx-\overline{y})$ and additively
regularized by von Neumann-Tichonov manner. The space $\R^n$ we
consider as a two different normed elementwise equal spaces
$\R^{n}_{\infty}$ and $\R^n_{2}$.

The space $\R^{n}_{\infty}$ is normed by $L_{\infty}$ norm
$\|a\|_{\infty}=\max\limits_{i}|a_{i}|$,
$(a=[a_{1},a_{2},...,a_{n}]^{T}\in\R^{n})$ and $\R^n_{2}$ space is
normed respectively by $L_{2}$ norm
$\|a\|_{2}=(\sum_{i=1}^{n}a_{i}^{2})^{\frac{1}{2}}$.

The corresponding matrix spaces $[\R^{n}_{\infty},\R^{n}_{\infty}]$
and $[\R^n_{2},\R^n_{2}]$ are normed respectively by
$\|A\|_{\infty}=\max\limits_{i}\sum_{j=1}^{n}|a_{ij}|$,
$A=\{a_{i,j}\}\in[\R^{n}_{\infty},\R^{n}_{\infty}]$ and by
$\|A\|_{2}=\lambda_{m}$, where $\lambda_{m}$ is the maximal spectral
number of the matrix A \cite{Col64}. Note, the norm $\|A\|_{\infty}$
is well defined. It is simultaneously completed and consisted to the
vector norm $\|a\|_{\infty}$.

The iterator (5) produces two different approximating sequences
subjects to the used regularizators.

To solve an ill-conditioned  \cite{Tih79} problem (4) we use the
process (5) in the $\R^{n}_{\infty}$ space with autoregularizator of
the process \cite{Ale70, Ale71}

\begin{equation}
\epsilon^{k}=\h\left( \sqrt{(\t^{k})^2+4N\r^{k}} -\t^{k}\right),
\end{equation}
where $\t^{k}=|| {f'}^T(x^{k}) f'(x^{k})   ||_\infty, \quad \r^{k}=
||{f'}^T(x^{k})(fx^{k}-\overline{y})||_\infty$ and $N=const>0$. The
behavior of the autoregularizator is accordant semilocal convergence
\cite{Ale71} of the process (5). As a rule the convergence domain
expands with increasing the initial value of the regularizator
$\varepsilon^{0}$.

To solve a singular problem \cite{Tih79}(4), we use the process (5)
in the space $\R^n_{2}$ with autoregularizator (6) with

\begin{equation}
\tau^{k}=\lambda^{k}_{m}
\end{equation}

is the smallest non-negative  eigenvalue of the matrix
${f'}^T(x^{k}) f'(x^{k})$ and

\begin{equation}
\r^{k}=
||{f'}^T(x^{k})(fx^{k}-\overline{y})-\alpha{^k}\overline{x})||_2,\;
\alpha{^k}=\gamma exp(-\sigma(k-1)),\;\gamma,\sigma = {\it
const}\geq 0.
\end{equation}

Presently in the case of singular problem we have not good semilocal
theory for the convergence of the process (5)-(8). In the case of
when convergence persists to a point $x^{*}$, the value
$\|x^{*}-\overline{x}\|_{2}$ is minimal. Usually one takes
$\overline{x}=0$. However the proposed iterator is fully applicable
and was used for analysis of large diversity of problems.

The described main iteration procedure starting from
Levenberg-Marquardt years till present days was applied extremely
successful.

\subsection{From local root extractors to global iterative processes}

In order to find all solutions of Eq.~(\ref{vector}) in the domain
$\D_f $ the vector $Fx^k$ is repeatedly multiplied by the
\textbf{local root extractor}
\[
e^{j}(x,\bar{x}^{(j)})= \frac{1}{1-\exp\left( - ||x - \bar{x}^{(j)}
||_2^2 \right)},
\]
in which $\bar{x}^{(j)} $ is the $j$-th solution of
Eq.~(\ref{vector}). In the repeated solutions of the transformed
problem \beq\label{F_M} F^{J} x:=\left( \prod_{j=1}^J
e^{j}(x,\bar{x}^{(j)}) \right) Fx=0, \quad J\geq 1, \eeq the process
(\ref{GN}) is executed with a new $Fx:=F^{J} x$. For every solution
the process (\ref{GN}) is started many times with different $x^{0}$
and $\epsilon^{0}$. Each time when $J$ increases the derivatives
$f'(x^{k})$ are computed analytically and the matrices
${f'}^T(x^{k})f'(x^{k})$  are adaptively scaled.

To analyze globally nonlinear problem (4) means to find all
solutions with estimation of themselves isolation in the domain
$\D_f $. For this purpose was build the mentioned above local root
extractor.

\section{The parameterization of ionization yield function Y }
The ionization yield function Y gives the number of ion pairs,
produced in 1 g of the ambient air at a given atmospheric depth by
one particle of the primary cosmic ray radiation with given kinetic
energy per nucleon.

The approximation is obtained following procedures similar to
described in \cite{Ale98,Ale99}. In this case we take the logarithm
of the problem, which permits to approximate large diversity of
distributions with similar shapes, having different amplitudes with
one model function with different parameters
\cite{Mis05a,Mis05b,Mis04}.

\subsection{Approximation of ionization yield function Y}
Usually the proposed approximation are carried out with large
diversity and classes of functions, some very complicated. It exist
powerful class of approximations - fractional-rational functions
applied in our case. As a result we obtain description of the
simulated data and fast convergence of the iteration process. For
example we utilize simple fractional-rational function

\begin{equation}
Y(h)=\frac{ah^{2}+bh+c}{dh^{2}+eh+f}
\end{equation}

where h is the atmospheric depth and a,b,c,d,e and f are parameters.
The fits for ionization yield function Y are presented in Fig.1-8
for protons with energies 500 MeV, 1 GeV, 5 GeV, 10 GeV, 50 GeV, 100
GeV, 500 GeV and 1 TeV. In the figures with solid black squares are
shown the obtained approximations and with open circles the
simulated data. During the solution of the inverse problem we take
logarithm of problem and scaling of the argument. This permits to
obtain parameterization with small uncertainties and fast
convergence of the process. In this case we solve ill-conditioned
problem with large number of condition $10^{17}$. The solution is
obtained after only 100 iterations steps. In this case the
normalized $\chi^{2}$ varies between 0.13 for 500 MeV protons and
0.001 for 1 TeV protons.

\begin{figure}[H]
\begin{center}
\epsfig{file=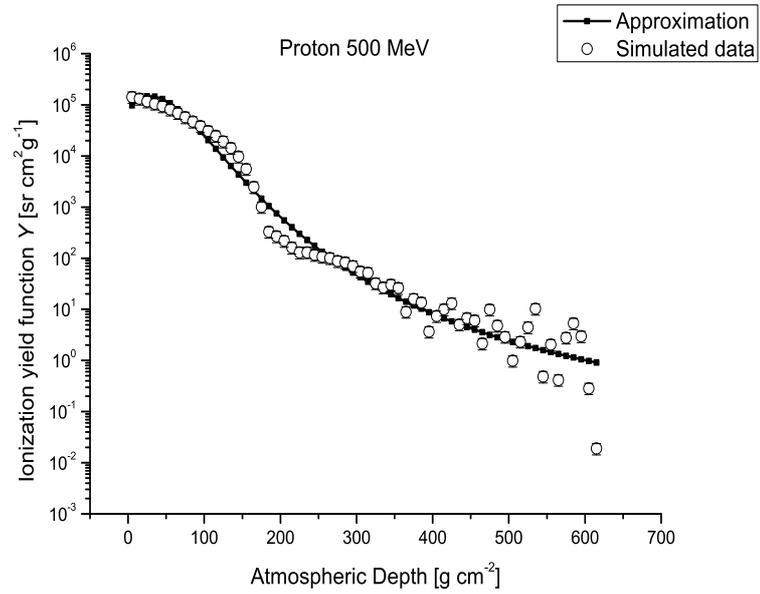,width=11cm, height=9cm} \caption{Approximation
of ionization yield function Y for 500 MeV primary protons}
\end{center}
\end{figure}

\begin{figure}[H]
\begin{center}
\epsfig{file=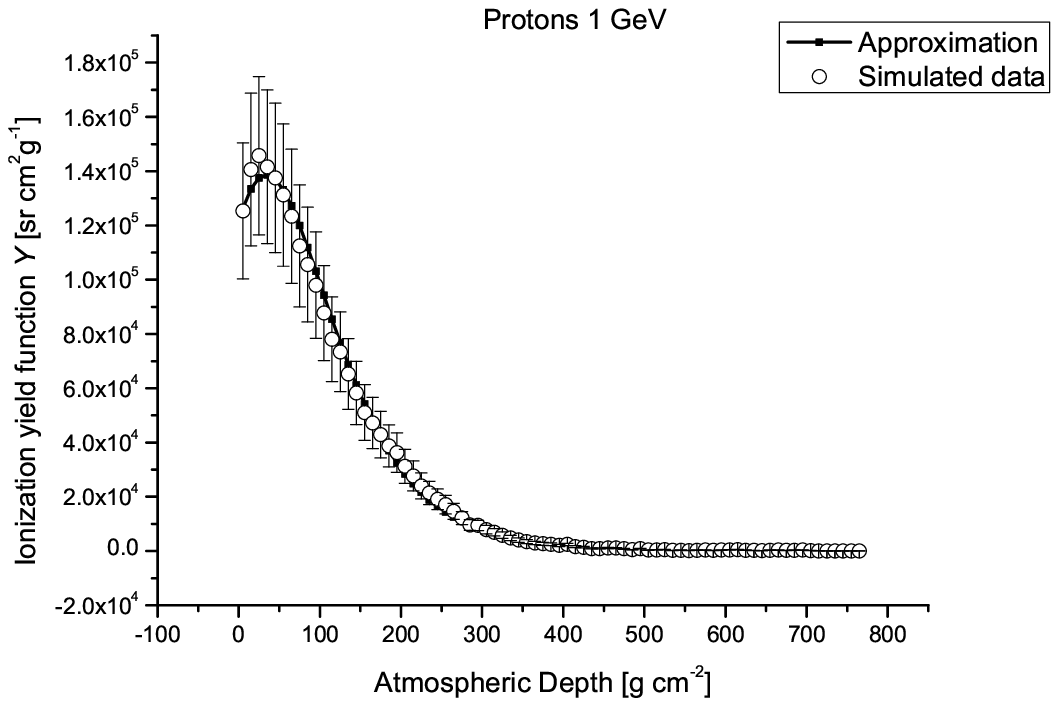,width=11cm,height=9cm} \caption{Approximation
of ionization yield function Y for 1 GeV primary protons}
\end{center}
\end{figure}

\begin{figure}[H]
\begin{center}
\epsfig{file=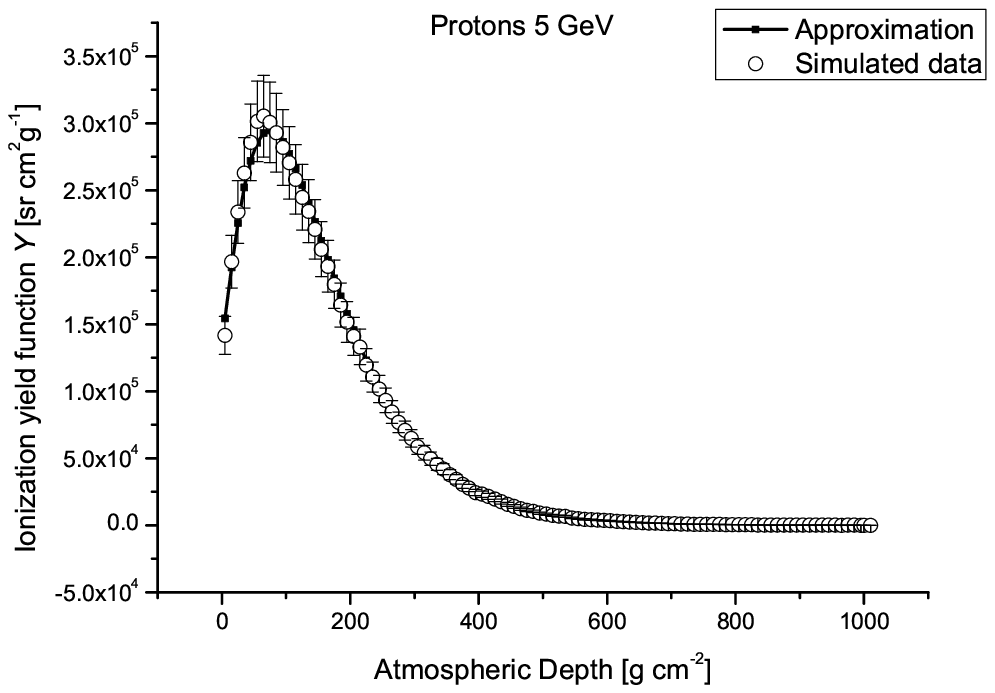,width=11cm, height=9cm} \caption{Approximation
of ionization yield function Y for 5 GeV primary protons}
\end{center}
\end{figure}

\begin{figure}[H]
\begin{center}
\epsfig{file=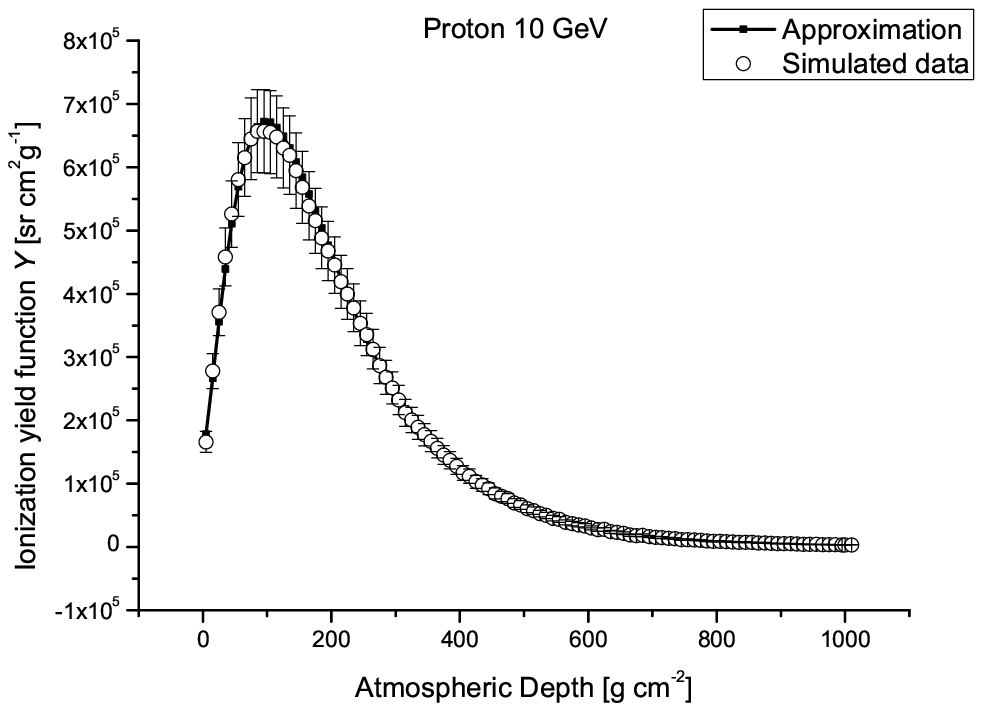,width=11cm,height=9cm} \caption{Approximation
of ionization yield function Y for 10 GeV primary protons}
\end{center}
\end{figure}

\begin{figure}[H]
\begin{center}
\epsfig{file=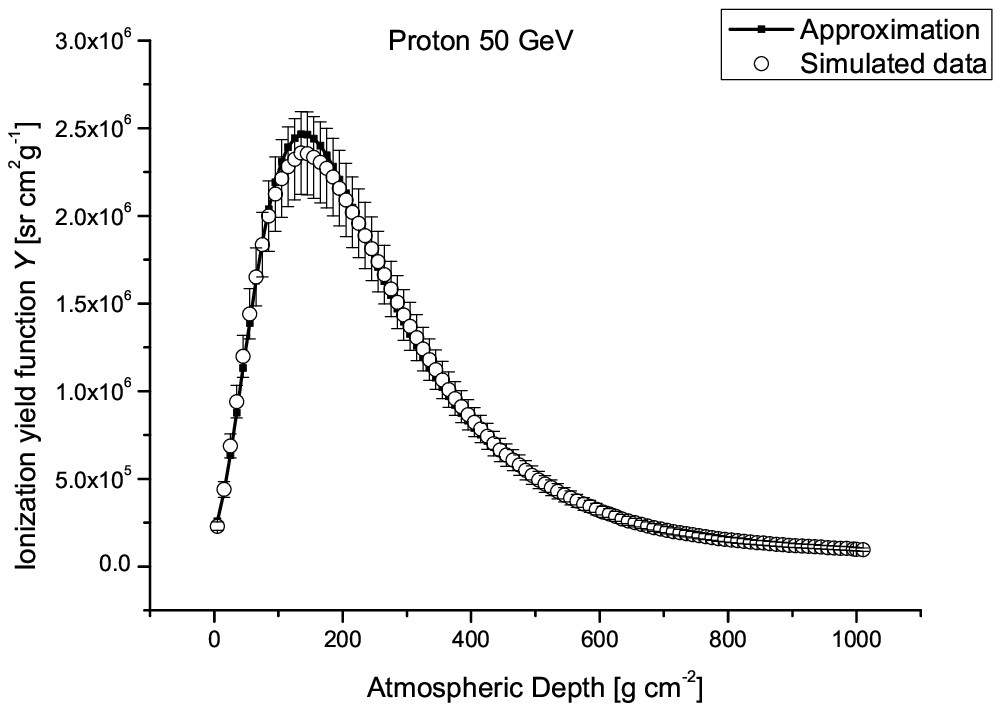,width=11cm, height=9cm} \caption{Approximation
of ionization yield function Y for 50 GeV primary protons}
\end{center}
\end{figure}

\begin{figure}[H]
\begin{center}
\epsfig{file=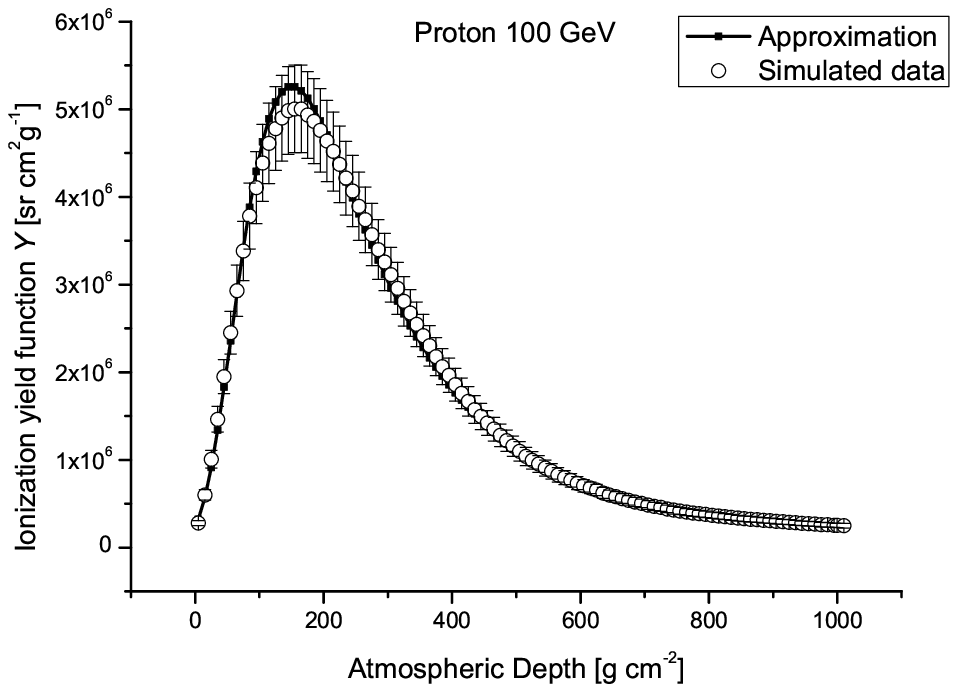,width=11cm,height=9cm} \caption{Approximation
of ionization yield function Y for 100 GeV primary protons}
\end{center}
\end{figure}

\begin{figure}[H]
\begin{center}
\epsfig{file=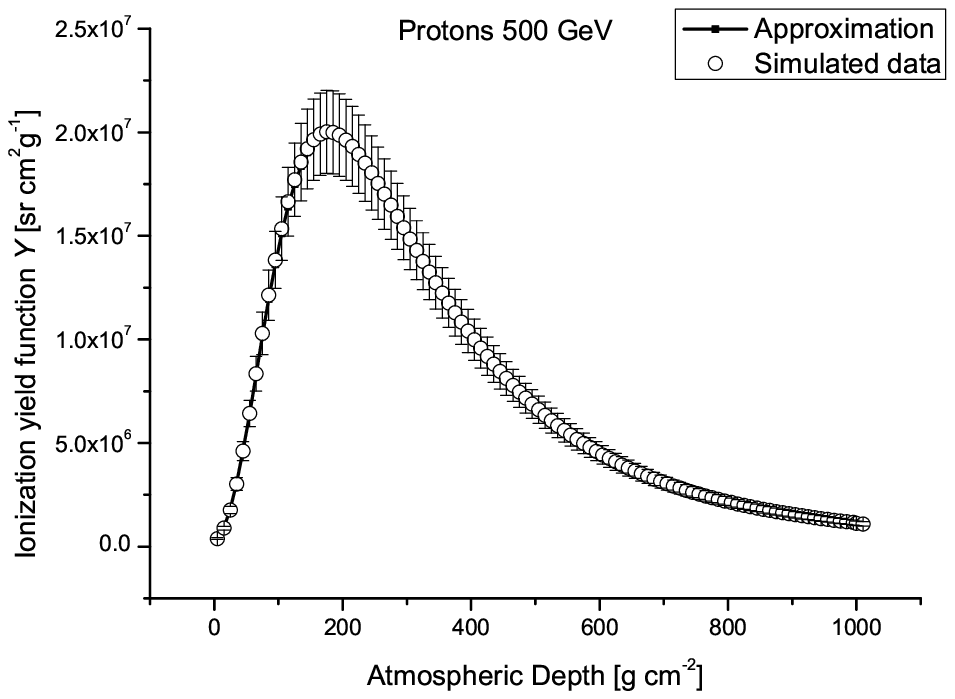,width=11cm, height=9cm} \caption{Approximation
of ionization yield function Y for 500 GeV primary protons}
\end{center}
\end{figure}

\begin{figure}[H]
\begin{center}
\epsfig{file=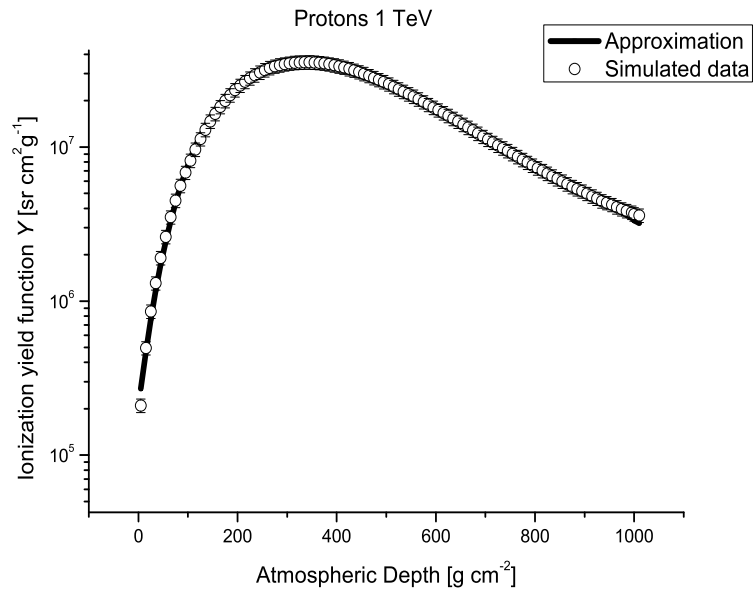,width=11cm,height=9cm} \caption{Approximation
of ionization yield function Y for 1 TeV primary protons}
\end{center}
\end{figure}

\subsection{Main results and parameters}
The quality of the parameterization is the same in all cases.

The general aim is in one hand the precise description of the
position of the Pfotzer maximum and on the other hand good
description in the low atmosphere.

 In the case of ionization yield
function for 500 MeV protons Fig.1 the end points of the
distribution are equally weighted. Our approximation is chosen to
take the mean values at distribution tail.

 In the remain cases the
main difficulties are connected with precise description of the
Pfotzer maximum. As example for the ionization yield function for 1
GeV and 5 GeV protons on observes very good coincidence between
approximation and simulated data (see Fig. 2 and Fig.3). The
approximated ionization yield function for 10 GeV protons  coincides
with simulated data (Fig.4).

In the case for 50 and 100 GeV, the parameterization  gives slight
increase of ion pairs in Pfotzer maximum (Fig. 5 and Fig.6).

Finally in the case of ionization yield function for 500 GeV and 1
TeV protons the approximation coincides once more with simulated
data (Fig. 7 and Fig.8).

 In all cases the
approximation in practice coincides with simulated data in all other
regions of the atmosphere,especially in the lower atmosphere. The
parameters of the approximation are presented in Table.1.

\begin{table}[htb]
\centering \caption{The parameters of the approximation}
\begin{tabular}{|c|c|c|c|c|c|c|}
  \hline
  Energy & a & b & c & d & e & f\\
  \hline\hline
  500 MeV & -6.09548 & 3.14499  & 0.31817 & 2.76023 & 0.37625 & 0.06494\\
  \hline\
  1 GeV & -2.51931 & 3.81687 & 0.78784 & 1.0136 & 0.64061 & 0.15493 \\
  \hline\
  5 GeV & -4.55444 & 6.69186 & 0.47114 & 0.2117 & 1.0687 & 0.09188 \\
  \hline\
  10 GeV & -0.89284 & 4.01468 & 0.19579 & 0.30881 & 0.5975& 0.03812 \\
  \hline\
  50 GeV & 1.82327 & 2.46817 & 0.15813 & 0.55512 & 0.31109 & 0.02995\\
  \hline\
  100 GeV & 2.28484 & 1.99474 & 0.13683 & 0.56878 & 0.22815 & 0.0255\\
  \hline\
  500 GeV & 0.99557 & 2.65916 & 0.10061 & 0.29285 & 0.30755 & 0.01891\\
  \hline\
  1 TeV & 0.03717 & 2.72562 & 0.24925 & 0.13949 & 0.27453 & 0.04701\\
  \hline
\end{tabular}
\end{table}

The approximation parameters are easy to fit with exponential or
power low functions. This permits using the expression for
ionization (3)  to estimate the cosmic ray induced ionization at
given location and altitude.

In addition, it is important that the approximation (10) is
analytically integrable , which gives the possibility to estimate
the total atmospheric ionization explicitly.

\section{Discussion}
The obtained parameterization of ionization yield function  Y
permits easily to compute the ionization rates from cosmic rays in
the Earth atmosphere for given location and conditions, instead of
using counting rates from different devices as a proxy. Moreover the
parameterization allows to estimate variations of the cosmic ray
induced ionization caused by the variable solar activity and as
result to evaluate several atmospheric effects of cosmic rays, on
different timescales and under different heliospheric conditions.

The cosmic rays may affect climate, and are probably important
climate driver. Recently, it was shown in \cite{Uso04b} that the
variations in the amount of cloud cover follow the expectations from
a cosmic-ray cloud cover link, specifically for low altitudes
\cite{Mar}. It was shown that the relative change in the low
altitude cloud cover is proportional to the relative change in the
solar-cycle induced atmospheric ionization at the given geomagnetic
latitudes. Namely, at higher latitudes the ionization variations are
about twice as large as those of low latitudes.

Above 100 percent saturation, the phase of water is liquid and it
will not be able to condense unless it has a surface to. Therefore
for formation of cloud droplets the air must have cloud condensation
nuclei. Changing the density of these particles, the properties of
the clouds can be varied. Having more cloud condensation nuclei, the
cloud droplets are more numerous but smaller, this tends to make
whiter and longer living    clouds. The suggested hypothesis, is
that in regions devoid of dust, the formation of cloud condensation
nuclei takes place from the growth of small aerosol clusters, and
that the formation of the latter is governed by the availability of
charge, such that charged aerosol clusters are more stable and can
grow while neutral clusters can more easily break apart.

If this process is dominant, charge and therefore cosmic ray
ionization would play an important role in the formation of cloud
condensation nuclei. With this in mind assuming our parameterization
(10) for ionization yield function Y and relation (3), as well the
parameterization for differential cosmic ray spectrum \cite{Buch} we
estimate the ionization rates for several cut-off rigidities
(Fig.9).

\begin{figure}[H]
\begin{center}
\epsfig{file=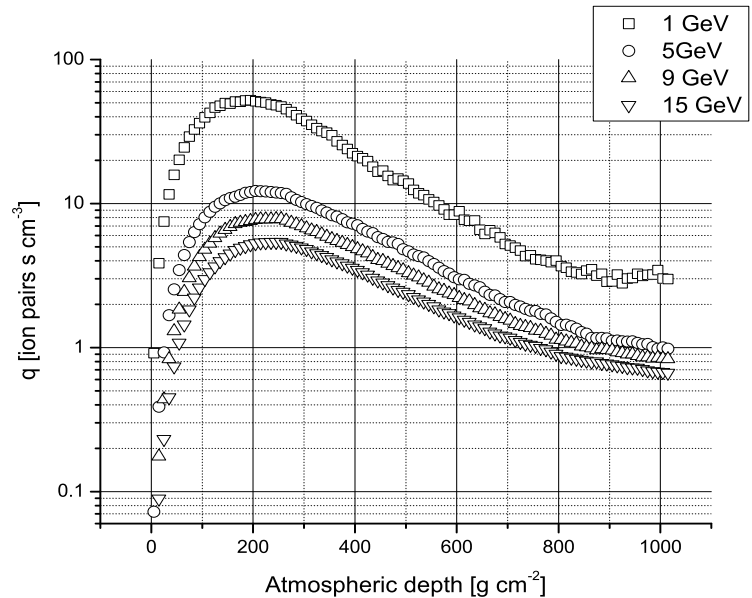,width=11cm,height=9cm} \caption{Ion pairs for
Proton induced showers for 1, 5, 9 and 15 GV rigidities}
\end{center}
\end{figure}

The presented results demonstrate realistic cosmic ray induced
ionization profile using average primary cosmic ray spectrum during
moderate solar activity. As it is seen these results are applicable
for different geomagnetic latitudes, respectively different
locations. The obtained results are in good agreement with the
profiles described in \cite{Uso06,Vel07}.

It exists the hypothesis that the intrinsic variation in the cosmic
ray flux are clearly evident in the geological paleoclimate data.
Within the determinations of the period and phase of the spiral-arm
climate connection, the astronomical determinations of the relative
velocity agree with the geological sedimentation record for when
Earth was in a hothouse or icehouse conditions \cite{Sha03}.
Moreover, it was found that the cosmic ray flux can be independently
reconstructed using the so called "exposure ages" of Iron meteorites
\cite{Sha02}.

 Thus it is very important, in one hand to investigate
the hypothetic influence of cosmic rays on the Earth's climate
trough the mechanism cosmic rays induced ionization-condensation
nuclei-cloud condensation nuclei-cloud cover, and on the other hand
to study the impact of intrinsic variation in the cosmic ray flux
during the motion of our planet in the Galaxy in attempt to estimate
and clarify their contribution.

\section{Conclusions}
In this work is presented new convenient parameterization of the
yield ionization function Y of primary cosmic rays into the Earth
atmosphere.  The parameterization was obtained on the basis of
inverse problem solution using afxy-code.

 The importance of cosmic
ray induced ionization is widely discussed, precisely the mechanism
related to clod cover formation.

Our results are important for precise estimation of the ionization
profiles in the atmosphere, when one deal with proton nuclei from
primary cosmic ray.
\\

\textbf{\Large Acknowledgments}\\
 The authors thank the Bogolubov
Laboratory of Theoretical Physics JINR, Dubna for hospitality and
support, especially Prof. V. B. Priezzhev. We acknowledge our
colleagues from Solar-Terrestrial Influences Laboratory - Bulgarian
Academy of Sciences.  We warmly acknowledge Prof. V. Yanke and Dr.
E. Eroshenko as well as Prof. L. Miroshnichenko from IZMIRAN Russia
for the fruitful discussions. Finally we thank  Dr. I. Usoskin from
Oulu university for several suggestions.


\end{document}